\documentclass[aps, prd, reprint, showpacs, superscriptaddress]{revtex4-1}
\usepackage{float}
\usepackage{microtype}
\usepackage{amsmath,amssymb}
\usepackage{linenoAMS}
\usepackage[usenames, dvipsnames, svgnames, table]{xcolor}
\usepackage[colorlinks=true, citecolor=RoyalBlue,
            linkcolor=BrickRed, urlcolor=ForestGreen]{hyperref}

\usepackage{graphicx}

\usepackage{siunitx}
\usepackage[capitalise]{cleveref}

\newcolumntype{d}[1]{D{.}{.}{#1}}
\newcolumntype{G}{>{\centering\arraybackslash}m{5em}}

\newcommand{\Tsrm}{T_\mathrm{SRM}}

\newcommand{\hEnergy}{\ensuremath{H}}

\newcommand{\parahead}[1]{{}}

\newcommand{\armPower}{\SI{750}{\kilo\watt}}
\newcommand{\sigBW}{\SI{450}{\Hz}}

\newcommand{\ifoSQZ}{\SI{12}{\decibel}}
\newcommand{\ifoLoss}{\SI{5}{\percent}}
\newcommand{\ROLoss}{\SI{10}{\percent}}
\newcommand{\SRCLoss}{\SI{0.1}{\percent}}

\begin{document}

\title{Tuning Advanced LIGO to kilohertz signals from neutron-star collisions}

\newcommand{\LIGOMIT}{LIGO Laboratory, Massachusetts Institute of Technology, 185 Albany Street, Cambridge MA 02139, USA}
\newcommand{\LIGOCaltech}{LIGO Laboratory, California Institute of Technology, 1200 E California Blvd, Pasadena CA 91125, USA}
\newcommand{\Birmingham}{School of Physics and Astronomy, and Institute of Gravitational Wave Astronomy, University of Birmingham, Edgbaston, Birmingham B15 2TT, United Kingdom}

\author{Dhruva Ganapathy}
\email[]{dhruva96@mit.edu}
\affiliation{\LIGOMIT}
\author{Lee McCuller}
\affiliation{\LIGOMIT}
\author{Jameson Graef Rollins}
\affiliation{\LIGOCaltech}
\author{Evan D. Hall}
\affiliation{\LIGOMIT}
\author{Lisa Barsotti}
\affiliation{\LIGOMIT}
\author{Matthew Evans}
\affiliation{\LIGOMIT}

\date{\today}
\begin{abstract}
Gravitational waves produced at kilohertz frequencies in the aftermath of a neutron star collision can shed light on the behavior of matter at extreme temperatures and densities that are inaccessible to laboratory experiments. Gravitational-wave interferometers are limited by quantum noise at these frequencies but can be tuned via their optical configuration to maximize the probability of post-merger signal detection. We compare two such tuning strategies to turn Advanced LIGO into a post-merger-focused instrument: first, a wideband tuning that enhances the instrument's signal-to-noise ratio 40--80\% broadly above \SI{1}{\kHz} relative to the baseline, with a modest sensitivity penalty at lower frequencies; second, a ``detuned'' configuration that provides even more enhancement than the wideband tuning, but over only a narrow frequency band and at the expense of substantially worse quantum noise performance elsewhere. 
With an optimistic accounting for instrument loss and uncertainty in post-merger parameters, the detuned instrument has a ${\lesssim}40\%$ sensitivity improvement compared to the wideband instrument.
\end{abstract}

\maketitle


\section{Introduction}
\parahead{Intro to BNS post-mergers:}
The discovery of a binary neutron star (BNS) merger by LIGO and Virgo in 2017
(GW170817)~\cite{Abbott_2017__1}, and the electromagnetic followup observations
of a kilonova~\cite{Abbott_2017__2}, have heralded a new era of observational
neutron star physics.  Information about the tidal deformability of the
constituent objects~\cite{Bernuzzi2012} is encoded in the gravitational
waveforms of binary mergers, both in the inspiral phase in the minutes before
coalescence~\cite{Hinderer_2010}, and in the post-merger phase promptly
after~\cite{Bauswein_2012}.

While the inspiral effects occur primarily below \SI{1}{\kHz}, the post-merger signal is expected at kilohertz frequencies~\cite{Clark2016, Chatziioannou2017}. Understanding the post-merger physics therefore requires improving or targeting detector sensitivity at these higher frequencies. In particular, post-merger waveforms have been simulated for various models of the neutron star equation of state (EoS) and their Fourier spectra typically show a narrow band of signal energy concentrated around \SI{2}{\kHz}~\cite{Bauswein_2012}.

\begin{figure}
	\centering
	\includegraphics[width=\hsize]{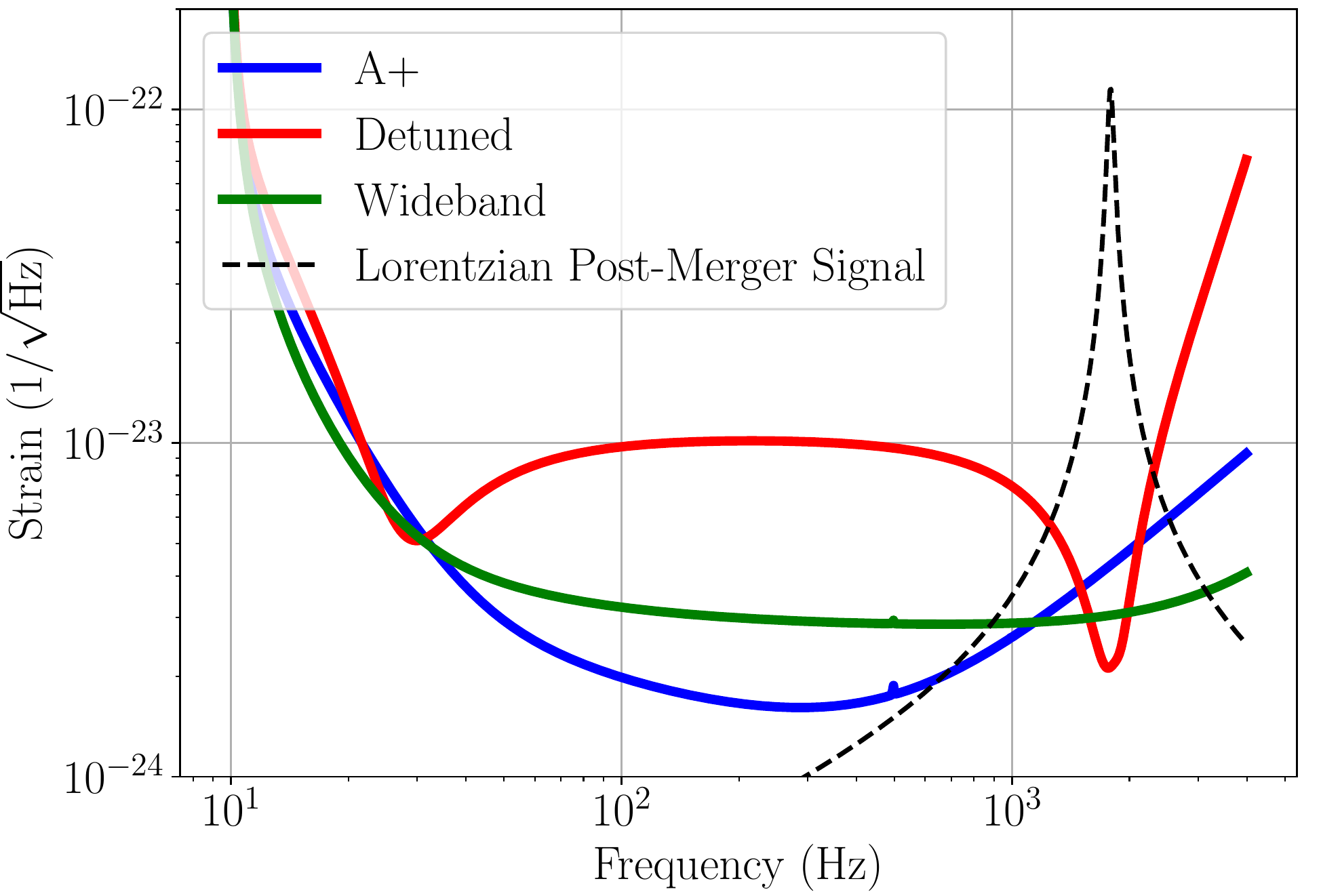}
	\caption{Interferometer configurations under comparison. Representative strain noise curves of the ``A+'', detuned, and wideband configurations are plotted for reference. Both altered configurations sacrifice sensitivity at low frequencies in order to increase high frequency sensitivity. The sensitivity improvement for the detuned configuration is across a relatively narrow band, and is achieved by detuning the signal recycling cavity in order to obtain a high frequency resonant enhancement. In the wideband configuration, the input transmission of the signal recycling cavity is reduced in order to increase the interferometer bandwidth.
    The dashed black curve corresponds to the strain $\tilde{h}_\text{DS}(f)$ of a lorentzian post-merger signal (\cref{eq:fd_lr_sym}) with $f_0 = \SI{1798}{\Hz}$ and $Q = 28.32$, observed at \SI{50}{Mpc} with $0.1M_\odot$ converted to gravitational wave energy during post merger (see~\cref{eq:energy}). It is plotted in spectral density units using the form $2\sqrt{f}|\tilde{h}_{\text{DS}}(f)|$.}
	\label{fig:strain_curves}
\end{figure}


\parahead{Concept of optimizing one or more detectors in a network and other work}
In a scenario where multiple gravitational-wave detectors are operational, it may be beneficial to maximize one or more detectors for sensitivity to these BNS post-merger signals, while relying on other detectors in the network for inspiral detection and source localization. Optimizing detectors at high frequencies has been investigated in the context of future major upgrades in current and new facilities~\cite{Miao_2018,Martynov_2019}, and in a proposal for a new dedicated high-frequency gravitational-wave interferometer~\cite{NEMO}. Here, we quantify the sensitivity to high-frequency, narrow-band post-merger signals for modified ``tunings'' of the LIGO interferometers and their upcoming ``A+'' upgrade~\cite{2015PhRvD..91f2005M,Aplus_design_curve}.

Two modifications are considered, with their strain spectra densities shown in Fig. \ref{fig:strain_curves}. The first is the ``wideband'' configuration, where the interferometer bandwidth is increased to encompass the expected post-merger resonances, and the second is the ``detuned'' configuration, where the A+ interferometer is operated with a high frequency, narrowband dip.
The only physical changes to the optical system associated with these new configurations are the transmissivity of the LIGO signal recycling mirror and filter cavity input mirror.
Neither of these new configurations requires modifying the facility, vacuum envelope or suspension design, so either could be readily adopted as a near-term modification to an A+ LIGO interferometer.

\section{Interferometer Configurations}

\parahead{LIGO HF limited by quantum noise:}
The sensitivity of existing gravitational-wave interferometers at frequencies above a few hundred hertz is limited almost exclusively by quantum shot noise~\cite{O3instrument,Buonanno_2001__1}.
Quantum shot noise can be reduced by increasing power in the arms of the interferometer~\cite{Kimble2001}, injecting squeezed light into the output port~\cite{Tse_2019}, and by trading sensitivity at some frequencies for others
by changing the optical parameters of the interferometer\cite{Mizuno:1995iqa}.

\parahead{Introduction to SRC: }
The LIGO detectors~\cite{Buonanno_2001__1} use arm cavities to both increase the arm power and shape the interferometer frequency response, with the addition of a signal recycling mirror (SRM) to implement the ``resonant sideband extraction'' scheme~\cite{Mizuno1993}. The SRM forms a signal recycling cavity (SRC) that determines the detector bandwidth. In the baseline A+ configuration the SRC is operated to resonantly couple the signal out of the arm cavities, broadening the bandwidth of the detector from $\SI{40}{\Hz}$ to $\SI{450}{\Hz}$. The parameters considered for A+ are given in \cref{tab:A+}.

The parameters of the A+ design are optimized for detecting inspiral signals, with quantum noise and classical thermal noise similarly affecting the detection range. The detector bandwidth, adjusted by the SRM transmissivity, is chosen to balance the peak sensitivity determined by shot noise and the degradation at low frequencies caused by radiation pressure noise. 
Frequency-dependent squeezing~\cite{Kimble2001,2016PhRvL.116d1102O,2020PhRvL.124q1102M} is employed to enhance the interferometer sensitivity at all frequencies.

Representative strain noise curves for the wideband and detuned configurations described below are shown along with the A+ curve in~\cref{fig:strain_curves}.  

Notably, squeezing enhancement plays a crucial role when comparing these alternative configurations. At post-merger signal frequencies of ${\sim}\SI{2}{\kHz}$, squeezed vacuum states are temporarily stored in the signal recycling cavity, experiencing its round-trip loss, $\Lambda_{\text{SRC}}$, repeatedly over multiple traversals. For A+, this amounts to a loss of ${\sim} 10\Lambda_{\text{SRC}}$. 
The wideband and detuned configurations change the storage time of the signal recycling cavity, which can result in strongly degraded squeezing as the SRC loss becomes comparable to other loss in the system.

\begin{table}
\centering
\caption{Parameters of LIGO configurations}
\begin{ruledtabular}
\begin{tabular}{l|c|c|c}
Parameter & \multicolumn{3}{c}{Value} \\

\hline \hline
Arm power & \multicolumn{3}{c}{\armPower{}} \\
Classical noises & \multicolumn{3}{c}{Thermal noise \cite{Evans2013}} \\
SRC length &\multicolumn{3}{c}{55 m}\\
SRC loss ($\Lambda_{\text{SRC}}$) & \multicolumn{3}{c}{\SRCLoss }\\
 
\hline
Injected squeezing & \multicolumn{3}{c}{\ifoSQZ{}} \\
Injection loss & \multicolumn{3}{c}{\ifoLoss{}}\\
Readout loss &\multicolumn{3}{c}{ \ROLoss{}}\\
\hline
Filter cavity length &\multicolumn{3}{c}{300 m}\\
Filter cavity loss&\multicolumn{3}{c}{60 ppm} \\
\hline
 & A+ & Wideband&Detuned  \\
\hline
SRM transmission&0.325 & 0.05&\cref{tab:opt_params}  \\
SRC detuning&0$^\circ$&0$^\circ$&\cref{tab:opt_params}\\
Signal 3dB bandwidth & \sigBW & 4.8 kHz & Fig. \ref{fig:SNR_comp1}  \\
Filter cavity transmission& 0.0012 &0.0004& Table \ref{tab:opt_params}  \\ 
Filter cavity detuning& 46 Hz &16 Hz&  Table \ref{tab:opt_params} \\ 
\end{tabular}
\end{ruledtabular}
\label{tab:A+}
\end{table}

\subsection{Wideband} 

The wideband configuration increases the bandwidth of a LIGO interferometer by further reducing the SRM transmissivity.
We consider $\Tsrm = 0.05$, reducing the peak strain sensitivity, but extending the bandwidth beyond 3 kHz. This value is chosen so
that the interferometer is sensitive to a wide range of frequencies where BNS post-merger signals are expected to lie.
This configuration is not optimized for any particular post-merger model, so it is effective in determining uncertain signals amongst potential models.

The decrease in peak sensitivity additionally reduces quantum radiation-pressure noise, requiring the filter cavity bandwidth to be decreased. This only affects sensitivity below \SI{100}{\Hz} and is not important for the analysis of post-merger signals. Similarly to A+, squeezing provides a broadband enhancement to the wideband configuration. Because the arms and SRC stay on resonance, the wideband configuration adds no additional frequency dependence to squeezing; however, decreasing $\Tsrm$ modifies how $\Lambda_{\text{SRC}}$ loss degrades the squeezing enhancement. 

In the wideband configuration, the loss added by the interferometer becomes ${\sim} 20\Lambda_{\text{SRC}}$ to ${\sim}40\Lambda_{\text{SRC}}$, increasing for signals approaching the detector bandwidth. The loss changes with frequency as the squeezing field transitions from being stored in the arms to being stored in the SRC, and the increased loss is due to the lower SRM transmissivity and correspondingly longer storage time. Even so, this increased loss is still subdominant to the input and output path losses, so squeezing performance is similar between the A+ and wideband configurations.

\subsection{Detuned}

The SRC can alternatively be operated in a ``detuned'' state, where it is held slightly off of resonance by
maintaining an optical phase shift using feedback control. In this state, the interferometer optical response forms a
resonant peak, resulting in a dip in the quantum noise spectrum in units of strain. This increases sensitivity at high
frequencies at the expense of sensitivity at lower frequencies~\cite{Buonanno_2001, Buonanno_2001__1, Buonanno_2002}.
When the detuning is optimized for resonances in the kilohertz region, an additional narrowband optomechanical spring
resonance is formed at low frequencies (\SIrange{10}{30}{\Hz}), but overall, this configuration is substantially less
sensitive for inspiral detection and localization at low frequency.

The detuned configuration is named for the microscopic ``detuning'' phase, $\phi_{\text{SRC}}$, added to the signal recycling
cavity. This phase moves the resonant frequency experienced by gravitational-wave signals. The
transmissivity of the SRM is additionally adjusted to narrow the resonance in the signal response and correspondingly deepen the dip in the noise spectrum. In the detuned configuration, $T_{\text{SRM}}$ and
$\phi_{\text{SRC}}$ must be optimized to achieve maximum signal-to-noise ratio (SNR) given a distribution of center
frequencies and signal bandwidths for postmerger signals. The configuration will depend on the particular post-merger model and the performance is computed for several parameter distributions which are described in~\cref{sec:appendix_distribution}.

Balanced homodyne readout is proposed for A+ as an improvement over LIGO's current fringe-offset readout \cite{Fritschel14}. For the detuned case, the interferometer signal sidebands are strongly imbalanced above and below the laser frequency at the resonant dip, so there is not a preferred readout angle for the postmerger signal detection. Varying the homodyne angle in interferometer models shows that the homodyne readout angle does not significantly impact the results or discussion for postmerger signals, and does not improve low frequency sensitivity beyond what's shown in \cref{fig:strain_curves}.

\parahead{Squeezing considerations for detuned operation: }
The detuned configuration considerably affects squeezing in two ways. First, the unbalanced optical response of the interferometer results in a frequency-dependent rotation of the squeezing quadrature, which must be compensated using a similarly unbalanced filter cavity. Along with the SRC parameters, this analysis optimizes the filter cavity input mirror transmission and resonance frequency $\Delta\omega_{\text{FC}}$ to maximize average SNR (see \cref{eq:eta}) for each parameter model. The filter cavity round trip loss is kept constant at the expected A+ level of \SI{60}{ppm}, which is close to the value already achieved at TAMA~\cite{Capocasa2018}.

Second, for the parameters of \cref{tab:A+}, the lower SRM transmissivity required for detuning causes the interferometer to inflict a squeezing loss of ${\sim}200\Lambda_{\text{SRC}}$ within the narrow frequency band of the optical resonance. This loss is equal or greater than the expected total input and output losses, which prevents squeezing from providing as large a benefit to the peak strain sensitivity in the detuned case as for the wideband or A+ cases, even with the optimized filter cavity parameters. Instead, the frequency dependence of the effective loss outside the interferometer bandwidth causes the squeezing to increase the effective band of the dip in strain spectral noise density.

\begin{table}[b]
    \begin{ruledtabular}
    \begin{tabular}{ c c c c c c c c}
        \textbf{Distribution}&$T_{\text{SRM}}$ &$\phi_{\text{SRC}}$&$T_{\text{FC}}$&$\Delta\omega_{\text{FC}}$ (Hz)&$\eta$&$\eta_{{\text{WB}/\text{A+}}}$\\
        \hline
        LQLF&0.72\%&2.38$^\circ$&0.0030&1843&1.29&1.38\\
        MQLF&0.70\%&2.36$^\circ$&0.0029&1846&1.34&1.41\\
        HQLF&0.69\%&2.36$^\circ$&0.0028&1850&1.37&1.42\\
        LQMF&0.83\%&1.37$^\circ$&0.0062&2766&1.04&1.76\\
        MQMF&0.78\%&1.37$^\circ$&0.0059&2765&1.07&1.81\\
        HQMF&0.76\%&1.36$^\circ$&0.0056&2764&1.09&1.83\\
        HQHF&1.13\%&0.76$^\circ$&0.0104&3783&1.00&2.10\\
    \end{tabular}
    \end{ruledtabular}
    \caption{Optimal interferometer configurations and improvement factors ($\eta$) for various astrophysical distributions, which are described in~\cref{tab:distributions}. The SRC is \SI{55}{\meter} long with a roundtrip loss $\Lambda_{\text{SRC}}$ of \SRCLoss. The filter cavity is \SI{300}{\meter} long with a roundtrip loss of \SI{60}{ppm}. The last column $\eta_{{\text{WB}/\text{A+}}}$ shows the average SNR improvement provided by the wideband configuration with respect to A+}
    \label{tab:opt_params}
\end{table}

\begin{figure*}
	\centering
	\includegraphics[width=\hsize]{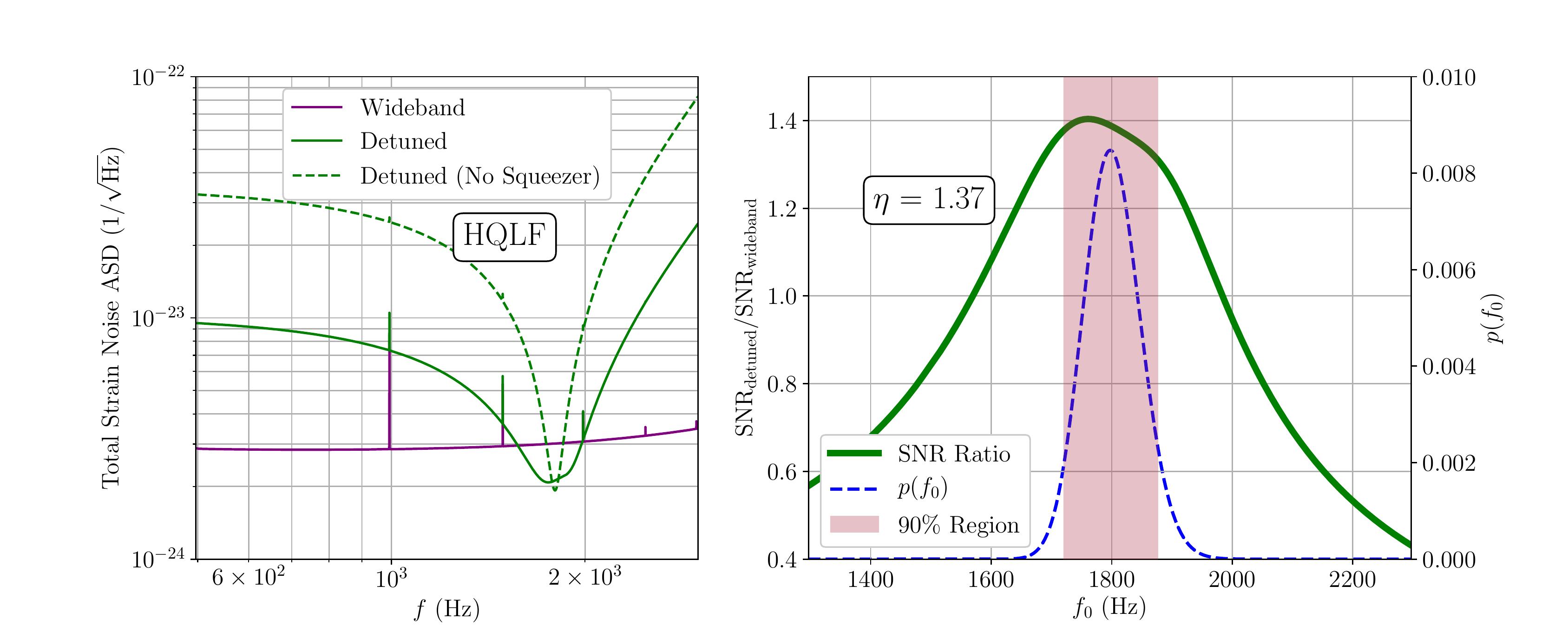}
	\caption{Performance comparison between the detuned and wideband configurations for the high-Q-low-frequency (HQLF) distribution of BNS post-merger signals (\cref{tab:distributions}) which peaks at $f_0 = \SI{1.8}{\kHz}$, and has $Q = 60$. The left plot shows the strain noise curves for wideband, detuned and purely detuned (without squeezing) configurations. For the detuned configurations, the enhancement provided by squeezing in the resonant dip is degraded due to the roundtrip loss of the SRC. However, the width of the dip is broadened significantly. The plot on the right shows the SNR ratio (green trace) of the detuned configuration with respect to the wideband configuration over a range of lorentzian central frequencies. The dashed blue trace corresponds to the probability distribution of signals as a function of central frequency. The detuned interferometer has been optimized to maxmimize $\eta$ for the probability distribution HQLF with $Q=60$ and $f_0 $ centered at $1.8$ kHz. The parameters of the detuned configuration are given in the third row of~\cref{tab:opt_params}. The shaded magenta region corresponds to the region containing 90\% of the signal probability. The overall SNR improvement $\eta$ is calculated to be 1.37 for this configuration. 
	}
	\label{fig:SNR_comp1}
\end{figure*}

\section{Binary Neutron-Star Post-merger Templates}
\label{sec:FOMS}

Binary neutron star post-merger waveform models consistently show that much of their gravitational strain signal energy is contained within a limited frequency band~\cite{Bauswein_2012}. For the purpose of comparing the SNR of detections, we approximate each post-merger narrowband signal as a damped sinusoid~\cite{Tsang2019}, using the subscript DS, which has a frequency-domain representation that is the symmetric composition of positive and negative frequency complex Lorentzian damped envelopes, using the subscript DE:
\begin{align}
    \tilde{h}_\text{DS}(f)
    &= \tilde{h}_{\text{DE}}(f)+\tilde{h}_{\text{DE}}^*(-f),
    \label{eq:fd_lr_sym}
      \\
     \tilde{h}_\text{DE}(f)
     &= \sqrt{\frac{\hEnergy}{4\pi}}\cdot\frac{e^{i\theta(f)} \sqrt{{f_0/Q}} }{f_0/2Q + i(f-f_0)}.
     \label{eq:fd_lr}
 \end{align}
$f_0$ is the signal's central frequency and its bandwidth is set by its $Q$-factor.  $H$ is the total energy of the strain signal. $e^{i\theta(f)}$ indicates additional parameters in the phase response~\cite{Tsang2019}, but these do not affect SNR calculations using the interferometer power spectral density.
For the detuned configuration, the interferometer's optical resonance bandwidth and dip frequency produces the greatest SNR when it is well matched to the waveform bandwidth and center frequency, but due to loss, the interferometer dip is generally of lower $Q$ than the templates.

The ability to match the detector to the signals is limited by the natural variability in the center frequency of post merger waveforms. Simulations of neutron star inspiral models have informed phenomenological relations between astrophysical system parameters and parameters of the resulting post-merger signal~\cite{Bernuzzi2015,Tsang2019}. These relations lead to a varying waveforms with distribution function $p(f_0)$, resulting from the distribution of binary neutron star systems. \cref{tab:opt_params} shows seven such distributions and their associated optimized interferometer parameters. These distributions, derived in \cref{sec:appendix_distribution}, are not tied to specific neutron star physical models, but instead span the uncertainty of the phenomenological waveform parametrizations.

\parahead{Introduction of improvement factor: }
The SNR is calculated from each configuration's power spectral density (PSD) and each distribution's waveform templates, $\tilde{h}_{\text{DS}}(f)$, using
\begin{align}
  \text{SNR}^2_{\text{config}} &= 4 \int_0^{\infty} \frac{|\tilde{h}_{\text{DS}}(f)|^2}{\text{PSD}_{\text{config}}(f)}df.
\end{align}
The average ratio of the detuned SNR to the wideband SNR, weighted over the distribution $p(f_0, Q)$, provides a figure of merit, $\eta$, to compare configurations:
\begin{equation}
    \eta^2 \equiv \int p(f_0, Q)
        \frac{\text{SNR}^2_{\text{detuned}}}{\text{SNR}^2_{\text{wideband}}}\,df_0\,dQ.
    \label{eq:eta}
\end{equation}
Similarly, the efficacy of the wideband configuration compared to the baseline of A+ is expressed as
\begin{equation}
    \eta^2_{{\text{WB}/\text{A+}}} \equiv \int p(f_0, Q)
        \frac{\text{SNR}^2_{\text{wideband}}}{\text{SNR}^2_{\text{A+}}}\,df_0\,dQ.
    \label{eq:eta_wb}
\end{equation}

\begin{figure*}
	\centering
	\includegraphics[width=\hsize]{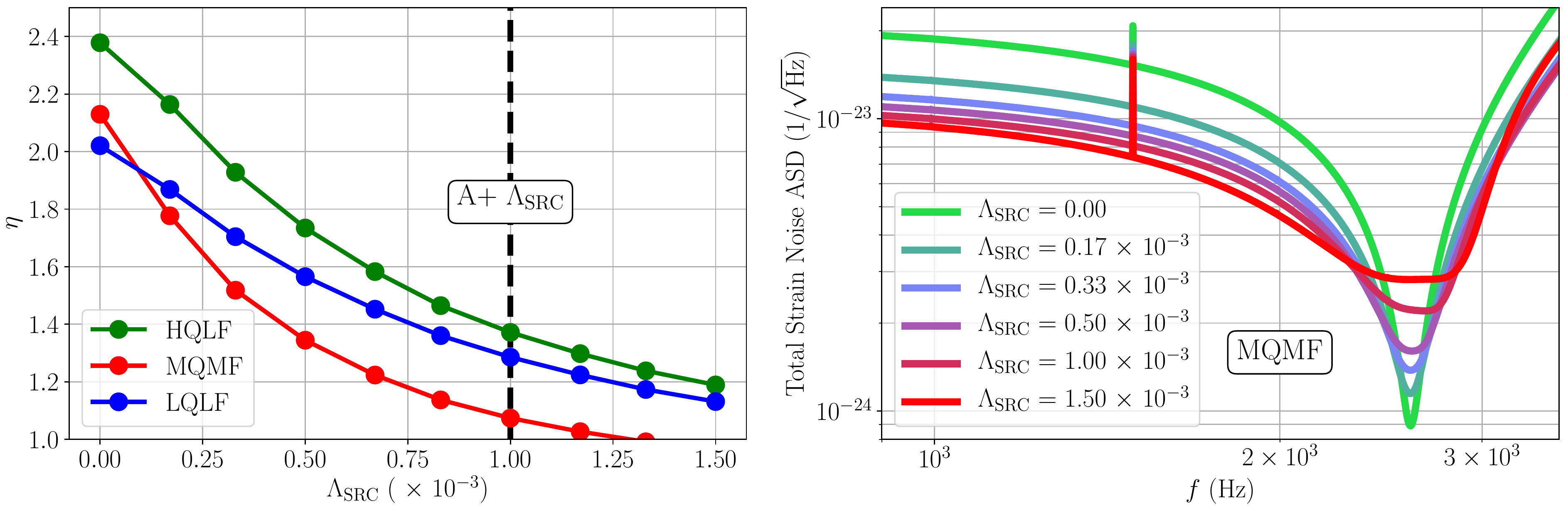}
	\caption{
    The effect of SRC loss on BNS post-merger sensitivity. The plot on the left demonstrates how the relative improvement factor for optimized interferometers (\cref{eq:eta}) is limited by the round trip SRC loss $\Lambda_{\text{SRC}}$ for various distribution models used for interferometer optimization. The A+ SRC loss limits sensitivity improvement to around 50\% of the zero loss case. The right plot shows optimum strain curves for distribution model MQMF as SRC loss $\Lambda_{\text{SRC}}$ is varied. Loss lowers the Q of the resonant band and squeezing widens the band to create a flat response. To maintain optimal performance, the resonance and squeezing effects of loss on dip bandwidth are balanced using the SRM transmission.
  }
	\label{fig:SRC_loss}
\end{figure*}

\section{Results}

\cref{tab:opt_params} shows the optimal interferometer parameters and relative improvement that is achieved by detuning the SRC for various astrophysical distributions of BNS post-merger signals. The wideband configuration provides an average SNR improvement $ \eta_{{\text{WB}/\text{A+}}}$ of $1.38-2.10$ over A+. The average SNR improvement from detuning, $\eta$, lies between 1.00 and 1.37 relative to the wideband for the optical parameters of \cref{tab:A+}.

The improvement provided by detuning the interferometer is generally lower for distributions that center around higher frequencies and lower $Q$ factors. The best case, corresponding to the distribution labelled HQLF is presented in~\cref{fig:SNR_comp1}, which shows strain noises and the SNR improvement over a range of lorentzian signals.  This configuration has an $\eta = 1.37$.

From the shaded magenta region in~\cref{fig:SNR_comp1}, that encloses 90\% of the signals, the detector dip is sufficiently wide to benefit the entire range of expected parameters. This is notable as the astronomical distribution of the center frequencies has a spread of approximately \SI{200}{\Hz}, which makes the distribution cover a wider band than the templates themselves for $Q > 10$. Because of this spread, even if the interferometer were lower loss and could obtain higher peak strain sensitivity in the detuned configuration, it cannot be configured to optimally match the interferometer resonance to the template resonance due to the astrophysical distribution of center frequencies. Instead, loss widens the sensitive band to naturally cover the distribution of templates, but reduces the peak sensitivity and relative SNR improvement.

\parahead{What about model uncertainty?: }
The relative benefit for the detuned interferometer, reported independently for each model distribution, provides a best case analysis where the astrophysical model of BNS post-mergers is assumed to be sufficiently constrained to optimize for specific parameters. The relative SNR can be cubed to represent the relative improvement to the detection volume or, relatedly, the relative rate of detections. The largest $\eta = 1.37$ corresponds to a factor of 2.57 in detection rate over the wideband configuration. If the post-merger model is not known, the detuning center frequency must be scanned, distributing time amongst potential detection frequencies. Scanning can thus significantly reduce the rate benefit of detuning. The wideband optimization has improved sensitivity at all of the potential models and avoids the need for scanning.

\section{Additional considerations}

\subsection{Loss in the signal recycling cavity}

\cref{fig:SRC_loss} shows the effect of SRC loss on the relative improvement, $\eta$. This figure indicates how severely the SRC detuning method is limited by optical loss within the SRC. The decreased SRM transmissivity required for a narrow band response creates an optical cavity where signal field crosses several optic surfaces and substrates such as the beamsplitter many more times than in the A+ or wideband configurations. Because of its use of optical resonance, the detuned configuration can nearly saturate the sensitivity available given the loss~\cite{Miao2019}, and squeezing tends to simply increase the bandwidth at peak sensitivity, as shown in the strain curve in~\cref{fig:SNR_comp1}. On the other hand, because of squeezing, the wideband configuration also approaches the maximum possible sensitivity, given loss, without sacrificing signal bandwidth.

A round trip power loss of $\SRCLoss$ is used for the interferometer models of this analysis. This value is optimistic and results from adding the expected losses across many wedged optics with anti-reflective coatings. Current measurements establish an upper bound of $0.3\%$ loss for the observing run 3 LIGO interferometers. This bound is derived from the bandwidth and optical gain measurements of the signal and power recycling systems, though for the same reasons that the A+ and wideband configurations have low loss sensitivity, this measurement also lacks sensitivity. The A+ upgrade intends to address issues such as beam clipping that impact loss, but is unlikely to drive the SRC loss below the optimistic value used, reiterating that the relative improvements quoted for detuning represent best-case scenarios.

\subsection{Operational Challenges}

Both the wideband and detuned configurations require lowering the SRM transmissivity from $\Tsrm \approx 30\%$, which will alter the operating parameters of the interferometer and require time to implement. In addition to the signal fields, the interferometers also employ radio-frequency fields to sense internal degrees of freedom related to the the power and signal recycling cavities, as well as the alignment of optics. The wideband configuration maintains the same operating modes for these fields and cavities, adjusted only by SRM transmission becoming $\Tsrm \rightarrow 5\%$.

The detuned configuration requires a more extreme adjustment with $\Tsrm \rightarrow 0.8\%$. In addition, detuned
operation results in imbalanced sidebands that not only impact the signal, but the RF control fields used for
alignment control and stabilizing internal degrees of freedom. Maintaining detuning using the current configuration of auxiliary fields requires adding control-point offsets, which can impact the reliability of continuous operation~\cite{Miyakawa2006,Hild_2007}. In total, detuning requires a considerable alteration of the operating controls and electronics, which would require investing observing time to implement.

\section{Conclusions}
\label{sec:conclusions}

Gravitational wave detectors are currently configured to maximize the detectability of binary inspirals, rather than of high frequency signals. This leads to the question of whether existing interferometers can be tuned to target science at high frequencies, where binary neutron star post-merger waveforms contain information about the equation of state of dense nuclear matter and its tidal deformability. This work analyzed two approaches: wideband and detuned configurations. The wideband configuration provides a relative SNR improvement in detecting postmerger spectral peaks of 40$-$110\% versus A+. 
 
Detuning LIGO's signal recycling cavity is another approach which maximizes sensitivity in a narrow band. For neutron star post-merger spectral peaks, detuning achieves an average SNR improvement between 0$-$40\% above the wideband configuration. Unlike the wideband configuration, the improvement from detuning is contingent on having empirical relations for the postmerger peak frequency and bandwidth as well as having the distributions of astrophysical parameters on which the relations depend. Both of the alternate configurations decrease the benefits imbued from quantum squeezing by increasing the influence of interferometer loss, but only the detuned configuration meaningfully diminishes squeezing. Detuning is also nontrivial to implement, and its relative SNR benefit does not appear sufficient to warrant trading implementation time with observing time, compared to the wideband configuration.

In the global network of gravitational-wave detectors, it could become advantageous to optimize one or more detectors for high frequency post-merger signals. This work suggests that adjusting the interferometer bandwidth, rather than detuning, is the most promising avenue while large uncertainties in the post-merger waveform exist.
 

\begin{acknowledgments}
The authors thank the National Science Foundation for support under grant PHY--0555406. EDH is supported by the
MathWorks, Inc. The authors acknowledge fruitful discussions with Haixing Miao and Hartmut Grote in the early stages of
this work, as well as valuable inputs from Stefan Ballmer, Teng Zhang and Peter Fritschel. Upper bounds on the SRC loss
were based on data from Valera Frolov.
\end{acknowledgments}

\appendix

\section{Astrophysical Distributions of BNS Post-Merger Lorenztian Model Parameters}
\label{sec:appendix_distribution}

This section establishes the phenomenological parameterizations used for the waveform template distributions. The form for the templates and their underlying phenomenological fits is derived from a set of numerical binary neutron star inspiral simulations~\cite{Tsang2019}. The simulations and fits provide the general form for relating BNS system mass $M$ to the post-merger waveform central frequency, $f_0$. The cited work does not provide relations for the waveform $Q$, and this is discussed below. 

In \cref{sec:FOMS,} the BNS post-merger signal was modeled as a lorenztian with central frequency $f_0$ and quality factor $Q$. From ~\cref{eq:fd_lr}, the peak frequency-domain strain amplitude for the lorenztian is
\begin{align}
    h_{\text{peak-f}} &= |\tilde{h}_{\text{DS}}(f_0)|
    \approx 
    \sqrt{\frac{Q \hEnergy}{\pi f_0}},
    \label{eq:strain_peak_lr}
\end{align}
which may be related to the peak strain in the time-domain waveform as
\begin{align}
    h_{\text{peak-t}} &\approx \frac{2\pi f_0 }{Q} h_{\text{peak-f}}.
    \label{eq:strain_peak_t}
\end{align}
These peak strain formulas in time and frequency domains can be applied to Table I of \cite{Tsang2019} to derive the waveform $Q$ value for each numerical simulation.

Using the peak strain values and the $Q$, one can then determine the waveform signal energy, normalized by total mass and distance. The strain signal energy for a general template is
\begin{align}
    \hEnergy &= \int\limits_{-\infty}^{\infty} |h(t)|^2 dt = \int\limits_{-\infty}^{\infty} |\tilde{h}(f)|^2 df
    \label{eq:strain_energy}
\end{align}
For signals of bandwidth small compared to the rate of change of the interferometer noise spectrum, this expression leads to $\text{SNR}^2 \approx 4 \hEnergy / \text{PSD}(f_0)$. This approximation is why strain signal energy provides a particularly morphology independent SNR metric to be computed from numerical simulations. Additionally, $\hEnergy$ can be related to the total energy emitted in the form of gravitational waves into the ringing post-merger signal. The energy in a strain signal is~\cite{Chatziioannou2017} 
\begin{align}
  E_{\text{GW}}
  &= \frac{c^3}{G}\frac{4}{5} \pi^2 D^2 \int\limits^{+\infty}_{-\infty}f^2 |\tilde{h}(f)|^2 \,df,
    \label{eq:energy_integral}
\end{align}
where $D$ is the distance to the source. This expression has an unphysical divergence if integrated to frequencies above $2f_0$ for the damped-sine model. When the integral is confined to frequencies where $h_{\text{DS}}$ is a good approximation, then in the limit $Q \gg 1$, the energy of a damped sine can be approximated as
\begin{equation}
	  E_{\text{GW}} = \frac{c^3}{G}\frac{4}{5} \pi^2 D^2 f_0^2 \hEnergy.
\end{equation}
$M_{\text{PM}}$ is the amount of mass that is converted to gravitational wave energy during the post-merger 
\begin{align}
    \hEnergy &= \left( 2 \pi f_0 D\right)^{-2} \frac{5 G}{c} M_{\text{PM}}.
    \label{eq:energy}
\end{align}


These waveform properties are used to formulate the dependence of the model templates on astrophysical parameters. The center frequency $f_0$ of the Lorentzian template model depends only on the total mass $M$ of the binary~\cite{Bernuzzi2015}:
\begin{equation}
	f_0(M,q) = \frac{C_1}{M}.
	\label{eq:f_M_q}
\end{equation}
The constant $C_1$ parameterizes this dependence, and it is related to the tidal deformability constant $\tilde{\Lambda}$ of the binary. The distribution of $f_0$ then depends on the astrophysical distribution of masses of neutron stars in merging binary systems. For this, we assume a gaussian distribution of neutron star masses~\cite{Ozel2012} centered around $1.35 M_\odot$ with a width of $0.05 M_\odot$.

\begin{figure}
	\centering
	\includegraphics[width=\columnwidth]{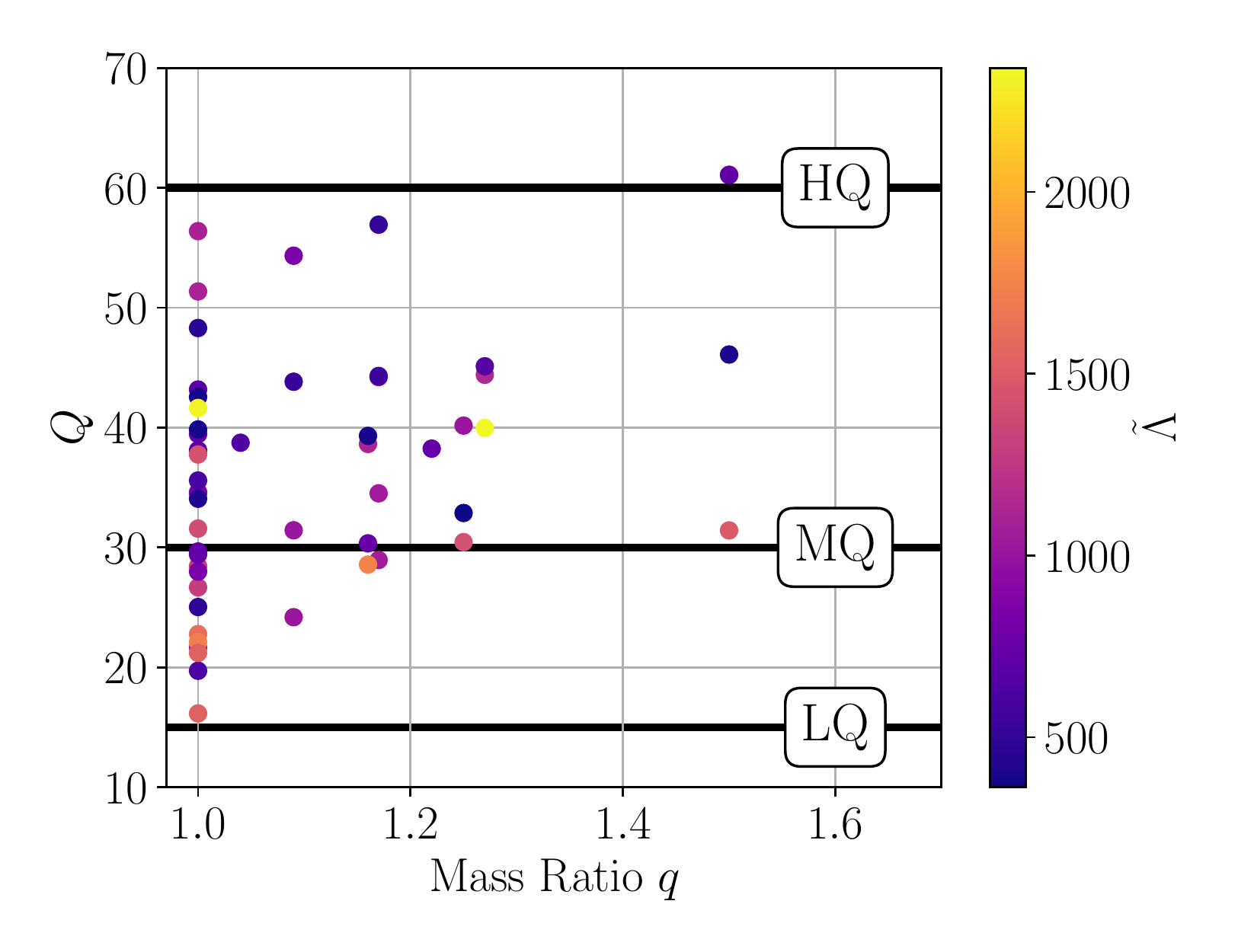}
	\caption{Lorentzian $Q$ factors inferred from simulation data contained in Table I of \cite{Tsang2019} plotted against the mass ratio $q$ of the binary. The colorbar shows the mass weighted tidal deformability $\tilde{\Lambda}$.  The solid lines coresspond to the values of $Q$ that have been chosen for the astrophysical distributions (see \cref{tab:distributions}) of Lorentzian signals in the analysis. These value are chosen in order to cover the range that is seen in simulation results.
	}
	\label{fig:scatter}
\end{figure}

In principle, the $Q$ is expected to depend on the binary's parameters, such as the mass ratio, the tidal deformability, and the equation of state. For a given post merger model, it could be assumed that the EoS and tidal deformability are constant, with the only important parameter varying astrophysically being the mass ratio $q$. Using \cref{eq:strain_peak_t}, Table I of \cite{Tsang2019} is used to calculate $Q$ and plot it against these parameters. \cref{fig:scatter} shows that the dependence of $Q$ on the mass ratio $q$ does not appear to follow any particular functional form, but $Q$ lies between 15 and 60 for mass ratios $q<1.6$. In order to remain within this range, $Q$ is unlikely to be a strong function of $q$. Because the expected mass ratio makes typical values of $q<1.2$, the astrophysical variability of $Q$ is sufficiently small that it does not affect this analysis, so it is fixed in each distribution. As a result, each distribution has only an astrophysical variation of $f_0$. The choice of constants, $C_1$ and $Q$ for each distribution is listed in \cref{tab:distributions}.

\begin{table}
    \begin{ruledtabular}
    \begin{tabular}{ c c c c }
        Distribution &$C_1$& $M(f_0)$ &$Q$\\
        &(kHz\,$M_\odot$)&(Hz)&\\
        \hline
        LQLF&4.86&1800&15\\
        MQLF&4.86&1800&30\\
        HQLF&4.86&1800&60\\
        LQMF&7.02&2600&15\\
       	MQMF&7.02&2600&30\\
        HQMF&7.02&2600&60\\
        HQHF&9.10&3333&60\\
    \end{tabular}
    \end{ruledtabular}
    \caption{Different astrophysical distributions for various choices of constants $C_1$in~\cref{eq:f_M_q}. The second last $M(f_0)$  represents the most likely $f_0$ for the distribution. Values of $C_1$ have been chosen in accordance with Table~1 of~\cite{Bernuzzi2015} to cover a range of frequencies (and tidal deformations). The distributions use fixed $Q$s which have been chosen to cover the range of values that are obtained from simulation results (See \cref{fig:scatter})} 
    \label{tab:distributions}
\end{table}

\newpage
\bibliography{main}

\end{document}